**Simple digital phase-measuring algorithm for low-noise heterodyne interferometry**

Wataru Kokuyama, Hideaki Nozato, Akihiro Ohta, and Koichiro Hattori

National Metrology Institute of Japan (NMIJ), National Institute of Advanced Industrial Science and Technology (AIST), Tsukuba, Ibaraki 305-8563, Japan

E-mail: *wataru.kokuyama@aist.go.jp*

**Abstract**

We present a digital algorithm for measuring the phase difference between two sinusoidal signals that combines the modified fringe-counting method with two-sample zero crossing to enable sequential signal processing. This technique can be applied to a phase meter for measuring dynamic phase differences with high resolution, particularly for heterodyne interferometry. The floor noise obtained from a demonstration with an electrical apparatus is $5 \times 10^{-8}$ rad/$\sqrt{\text{Hz}}$ at frequencies above approximately 0.1 Hz. In addition, by applying this method to a commercial heterodyne interferometer, the floor-noise level is confirmed to be $7 \times 10^{-14}$ m/$\sqrt{\text{Hz}}$ from 4 kHz to 1 MHz. We also confirm the validity of the algorithm by comparing its results with those from a standard homodyne interferometer for measuring shock-motion peak acceleration greater than 5000 m/s$^2$ and a 10 mm stroke.

# 1. Introduction

Heterodyne interferometry is a widely used and highly sensitive method for measuring displacements or optical-path variations and finds several applications, such as length metrology [1], measurement of intense shock [2], evaluation of micro-electromechanical devices [3], biomedical measurements [4], and space-based gravitational-wave detection [5]. The most important component for accurate and stable heterodyne interferometry is the digital phase meter [6], which calculates the phase difference between two beatnote signals. Recent applications for dynamic length measurements target a velocity greater than 1 m/s; thus, the modulation frequency exceeds several megahertz. However, conventional digital phase meters compromise between signal-frequency range and phase resolution. Therefore, an improved method is required that allows high-resolution displacement sensing in situations involving dynamic motion, such as vibration and shock.

Conventional digital phase-meter detection techniques are classified into three main types: lock-in, fringe counting, and zero crossing. The lock-in (or demodulation) technique is the most accurate and popular in many fields, including for making vibration measurements [7,8]. The signal to be measured is multiplied by the sinusoidal reference signals, which are normally generated by the digital processing unit. In this technique, the input-signal frequency must approximate the frequency of the reference signal because the demodulation process is equivalent to a single-bin discrete Fourier transformation (DFT). The lowest frequency of the DFT, or frequency resolution, corresponds to the inverse of the duration of the input signal subject to the DFT; therefore, the frequency range of the input signal is limited. For the same reason, the time constant of the lock-in amplifier (LIA) limits the frequency bandwidth of the signal that can be detected. In other words, heterodyne interferometry with a lock-in phase meter cannot be used to measure high velocities.

The fringe-counting technique has a wide dynamic range in velocity and has been a popular commercial heterodyne interferometry technique for measuring displacement since the 1970s [9]. In this technique, the circuit counts the number of fringes in the Doppler-modulated signal and in the reference signal. The counter takes the difference of these two numbers, and the phase difference between the two signals is a pi-multiple of the counter output. In principle, this technique offers limited resolution because the results are integer multiples of pi. Although some techniques have been proposed to enhance fringe counting [10–13], they demand extremely complex hardware.

The third type of digital phase meter detects the sampled signal zero-crossing time to estimate the signal frequency or its phase [14–18]. Many applications, including gas spectroscopy [19], power-management systems [20], and vibration metrology [21], have been investigated with such systems. However, owing to the limited computation power, none of them permit real-time processing of signals above several megahertz. In addition, no results can be obtained for constant time intervals, in contrast to homodyne interferometers [22,23], because the zero-crossing time of the modulated signal is not always in a constant period. Thus, these three conventional methods must be improved for making real-time measurements of dynamic motion with digital phase meters.

In the present study, we propose a simple algorithm for estimating digital phase measurements from sampled data. This algorithm essentially combines two conventional methods: the modified fringe-counting method and the two-sample zero-crossing method. The steps of the algorithm only involve simple additions and divisions; hence, real-time signal processing is possible, such as that required for field-programmable gate arrays (FPGAs). Contrary to the conventional zero-crossing method, results can be obtained for constant time intervals. Furthermore, we demonstrate the algorithm's potential by applying it to confirm the low noise levels imprinted in signals from a commercial laser vibrometer.

**2. Simple digital algorithm for measuring phase**

The principle of the proposed algorithm is graphically presented in Fig. 1. The phase $\phi(t)$ of the sinusoidal signal [Fig. 1(b)] linearly increases [Fig. 1(a)]. We assume here that the signal does not contain phase-fluctuation components higher than the signal frequency. To estimate the signal phase, we calculate the averaged phase $\frac{1}{k}\sum_{n=0}^{k-1}\phi(n)$, where $n$ is the data number and $k$ is the accumulation factor, which is set to 20 in this example. Graphically, this value is equivalent to the area under the line formed by the filled circles in Fig. 1(a). This area approximately corresponds to the area below the thick red line (using the edge at the zero-crossing time), which can be measured as follows: The area is divided in two by the narrow black line [formed by the triangles in Fig. 1(b)] and the rectangles, which is the residual between the thick line and the narrow line. For the first area, modified fringe counting is used as follows: (i) prepare a counter, which increments when the data crosses zero; (ii) integrate the data from the counter [the results represented by the triangles in Fig. 1(b)]. For the integration in step (ii), we use a linear approximation wherein we linearly interpolate between two points nearest to a zero crossing.

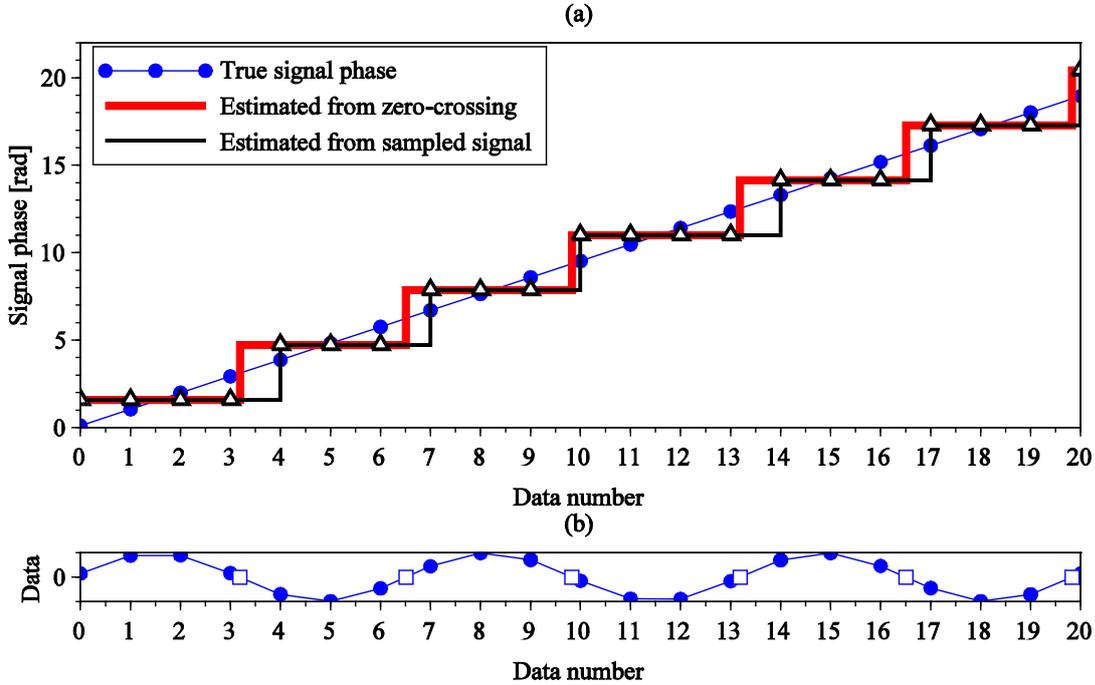

**Fig. 1.** (a) Principle of the algorithm. The signal's phase $\phi(t)$ is indicated by the filled circles. The thick red line and the narrow black line represent the signal phase estimated from the actual

zero-crossing time and from the sampled signal, respectively. (b) The sinusoidal input signal to the phase meter, which was sampled by an analog-to-digital converter. The actual zero-crossing time is indicated by the open squares.

This algorithm can be easily applied to real-time signal processing. In other words, the algorithm can process a serial signal using only two consecutive data points. Figure 2 shows the detailed procedure for a phase meter. In a phase meter, two logical blocks operate in parallel to measure the phase difference between two sinusoidal input signals. The phase meter comprises three functional blocks: (i) modified fringe counting, (ii) fraction processing, and (iii) accumulation.

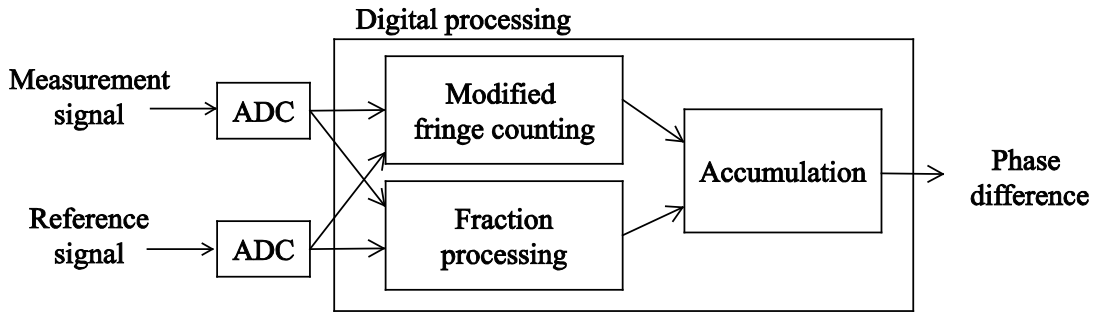

**Fig. 2.** Block diagram of overall signal-processing procedure for the proposed algorithm. ADC stands for analog-to-digital converter.

In the modified fringe-counting block, the count $C_i$ is incremented upon detecting the zero crossing of the measured signal and decremented upon detecting the zero crossing of the reference signal. If the two zero crossings are simultaneously detected, the count remains unchanged. Note that the value $C_i$ is generated at each clock step, which also drives the analog-to-digital converters (ADCs) and the digital signal-processing system.

Fraction processing should simultaneously run with the modified fringe-counting method. When a zero crossing is detected in the signal, the fraction-processing block calculates the average phase by linearly interpolating between two consecutive data points *A* and *B*, where *A* and *B* are the values just before and just after the zero crossing, respectively. The output is obtained as $F_j^X = \frac{|B|}{|A|+|B|}$. The suffix *X* refers to the signal type: measurement (M) or reference (R) signal. The suffix *j* corresponds to the index of the zero crossing.

The last part of the phase meter includes an accumulation block, which combines the output from the two previous blocks. In particular, the result $U$ is calculated as follows:

$$U = \frac{1}{N}\left(\sum_{i=1}^{N} C_i + \sum_{j=1}^{l} F_j^M - \sum_{k=1}^{m} F_k^R\right), \qquad (1)$$

where *N* is the preselected number of data points in the accumulation range. For example, when a 1-MHz-bandwidth result is desired and the ADC sampling rate is 500 MHz, *N* should be 500. In addition, *l* and *m* give the number of zero crossings and the range, respectively. *U* is the phase difference in a half-cycle unit and can be converted to the phase difference $\Delta\phi$ by using $\Delta\phi = \pi U$. This value can be used in posterior data processing. For instance, for single-path heterodyne interferometry, the displacement *d* is obtained by using $d = \frac{\Delta\phi}{2\pi}\lambda$, where $\lambda$ is the wavelength of the laser used in the interferometer.

## 3. Experimental demonstration

To test the proposed algorithm, we performed three experiments: a measurement of long-term intrinsic noise from an electronic setup, measurement of noise from a commercial laser Doppler vibrometer (LDV), and measuring shock motion to compare the results with those of standard homodyne laser interferometry.

Figure 3(a) shows a schematic of the apparatus used for measuring long-term electrical stability. An 80.3 kHz, $0.6\,\text{V}_\text{p-p}$ sinusoidal wave is generated by a signal generator (Tabor Electronics WS8251) and is input into channel 1 of a digitizer (National Instruments PXI-5922). The signal is split by using a T connector, and a 30 m cable is inserted into the signal path to create an artificial phase delay of about $1.2 \times 10^{-2} \times 2\pi$ rad. The delay lines comprise coaxial cables with polyethylene insulation. The split signal is sent to channel 2 of the digitizer. The sampling timing for both channels in the digitizer is locked to the rubidium frequency standard (Stanford Research FS725). The ADC sampling rate is set to 500 kHz, and the resolution is 24 bits. The sample size is 50,800,000, and the accumulation factor $N$ is set to 250. The coupling mode, input impedance, and input-voltage range for the digitizer is AC mode, 50 Ω, and $\pm 0.5$ V, respectively. According to the datasheet, the noise density of the digitizer is $-169$ dBFS/Hz (equivalent to $2 \times 10^{-9}$ V/$\sqrt{\text{Hz}}$ in this setup) at around 80 kHz.

We compare the results obtained with the digitizer with those obtained with a conventional phase-measurement system comprising a commercial LIA, as shown in Fig. 3(b). Instead of the digitizer in Fig. 3(a), the two-phase LIA (NF Corporation 5601B) detects the two sinusoidal signals. The time constant and input sensitivity of the LIA are set to 10 ms and 1 V, respectively. The in-phase and quadrature outputs, representing $A\cos\phi$ and $A\sin\phi$, are read by the same digitizer. Sampling rate, resolution, sample size, accumulation factor, and input impedance of the ADC is the same as the setup for Fig. 3(a). To acquire the signal from the LIA, the coupling mode and input-voltage range are switched to DC mode and $\pm 5$ V, respectively. The phase $\phi$ is reconstructed by taking the arctangent of the ratio of the in-phase output to the quadrature output.

We also make additional measurements with a low-pass filter (LPF, NF Corporation 3625) instead of the 30 m delay line to avoid deterioration in the LIA noise level for small phase differences. This setup is not shown in Fig. 3. The LPF cutoff is set to 120 kHz to add a sufficient phase delay of $\phi \sim 144°$ to one of the input signals. Any other setup is exactly the same as the measurement system for Fig. 3(b).

The DFT results are presented in Fig. 4(a). The thick black line shows the noise level for the proposed algorithm, whereas the thin black line shows the noise level for the comparative experiment with a 30 m delay line. The result with the LPF is the thin blue line. Above approximately 0.1 Hz, the floor noise of the proposed algorithm is $5 \times 10^{-8}$ rad/$\sqrt{\text{Hz}}$, whereas for frequencies below 0.1 Hz, the noise level is proportional to the inverse frequency ($f^{-1}$). This result shows that the noise level with the proposed algorithm is several orders of magnitude lower than that obtained with a conventional measurement using LIA, especially for low frequencies below 1 Hz, wherein digital signal processing benefits signal stability and noise level. This noise level is low enough to satisfy the requirements for phase meters in very precise scientific measurements, such as space-based gravitational-wave detectors [24].

(a)

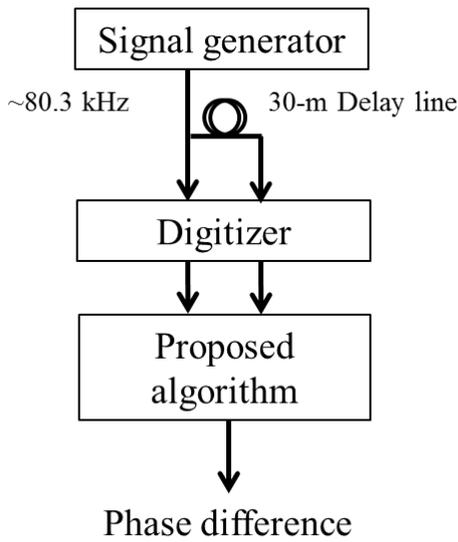

(b)

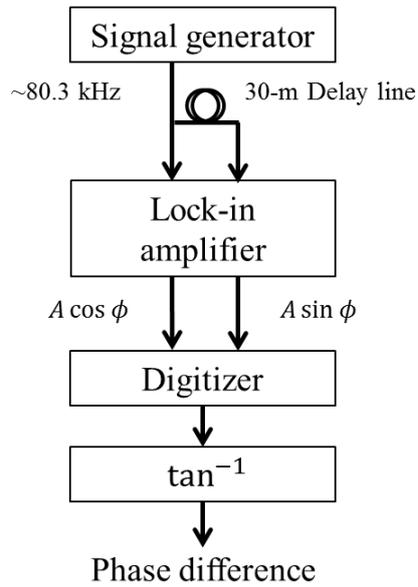

(c)

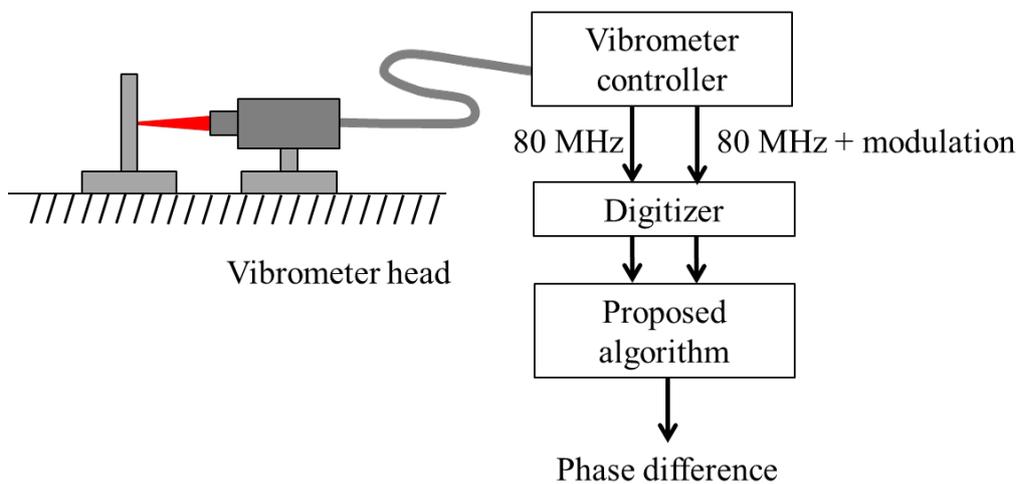

**Fig. 3.** Experiment for comparing noise-level measurements (a) by using only electronics and (b), (c) by using a commercial laser Doppler vibrometer.

Figure 3(c) shows the setup for measuring noise with an optical apparatus. The optical components of the laser, including optical fibers and lenses, acousto-optic modulator (AOM), photodetectors, and band-pass filters, are incorporated in the commercial LDV (Ono Sokki LV-1800), which includes a heterodyne interferometer with a modulation frequency of 80 MHz. Mounted on the vibrometer is an unstabilized He–Ne laser that, to comply with the Class-2 laser standard, emits less than 1 mW. A measurement laser beam is focused onto a fixed plane mirror. The reflected beam returns to the vibrometer head, and the interference fringes are detected. The Doppler-modulated signal and the reference signal are directly obtained from the vibrometer controller. Note that the reference signal, which is generated by a crystal oscillator inside the LDV, drives the AOM. The two signals are acquired by a digitizer (National Instruments PXI-5152) with a sampling rate of 500 MHz and with 8-bit resolution. Following this, to calculate the displacement, the proposed algorithm is applied in an offline analysis on a personal computer. To extend the bandwidth to 1 MHz, the accumulation factor *N* is set to 250. According to the recommendations of the Comité International des Poids et Mesures [25,26], the vacuum wavelength of the laser can be considered to be 632.9908 nm with a relative standard uncertainty of $1.5 \times 10^{-6}$, which is adequately accurate for the present measurements. Considering the empirical equation for the refractive index of air, we use a wavelength in air of 632.8 nm. The room temperature is maintained at 21.5 °C. Figure 4(b) shows the noise spectrum produced. The floor noise remains at $7 \times 10^{-14}$ m/$\sqrt{\text{Hz}}$ from 4 kHz to 1 MHz, indicating that the noise level is as low as for other laser interferometers [3,5].

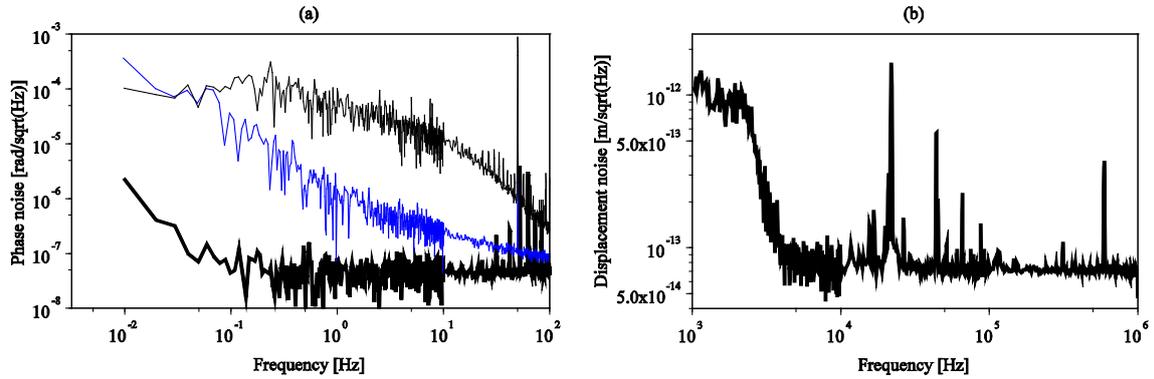

**Fig. 4.** (a) Results for the test with electronics. The thick line shows the result for the proposed algorithm. The thin black line shows the noise level with the lock-in amplifier and a 30 m delay line. The thin blue line shows for the result with the lock-in amplifier and an artificial phase delay implemented with a low-pass filter. (b) Results for the test with laser Doppler vibrometer.

In the third experiment, the LDV with the proposed algorithm is used to measure dynamic motion and the results are compared with those from a quadrature homodyne interferometer. We used a shock calibration system [27] based on the national shock acceleration standard. The system contains a shock exciter, which is able to generate a half-sine-shaped shock pulse in a peak-acceleration range between $200 \text{ m/s}^2$ and $10\,000 \text{ m/s}^2$. As a displacement-measurement standard, we made a quadrature homodyne interferometer based on a stabilized He–Ne laser (linewidth ~2 MHz). A pair of quadrature signals from the interferometer is detected by digitizer (National Instruments PXI-5152) at a sampling rate of 50 MHz and 8-bit resolution. The data are fit to elliptical Lissajous curves (a so-called Heydemann correction [28]) to reduce periodic nonlinearity. After this step, we apply the well-known phase unwrapping algorithm to obtain the displacement data. The LDV laser beam is aligned with that of the homodyne interferometer. The moving surface measured by the two interferometers is a 10-mm-diameter flat mirror attached on the anvil. A reference transducer is sandwiched by the anvil and the mirror, making no contribution to this experiment. The configuration of the LDV and other electrical apparatus for the LDV is the same as that for Fig. 3(c). More details of the other mechanical, optical, and electrical apparatuses are available in Ref. [27].

The results for simultaneous measurements of displacement are shown in Fig. 5(a) and the difference between the two displacements is presented in Fig. 5(b). From ~2.5 to ~3.0 ms, the shock motion is applied to the anvil. The calculated peak shock level is 5300 m/s$^2$, and the width of the half-sine-shaped shock pulse is about 0.5 ms. The velocity just after the impact is approximately 1.56 m/s. The two displacements are consistent within several tens of nanometers, showing a small mismatch at the subwavelength of the laser, even for a translation of over 10 mm. The difference is probably induced by a mechanical error comprising an alignment error between the two laser beams combined with an undesired parasitic motion of the aluminum anvil.

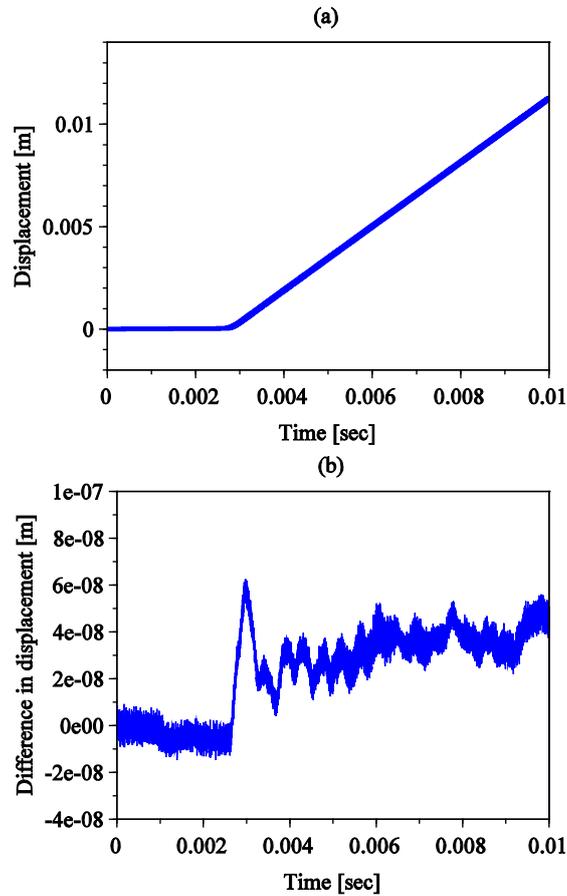

Fig. 5. (a) Measured displacement for half-sine-shaped shock of 5300 m/s$^2$. On this scale, the results of the LDV and those of the homodyne-interferometer measurements cannot be distinguished. (b) Difference between the two displacements shown in panel (a).

**4. Discussion**

The proposed algorithm combines high resolution with dynamic range: the resolution of the proposed algorithm exceeds that of fringe counting by processing only sampled data. If $n$ is the effective number of bits (ENOB) in the AD conversion and $N$ is the accumulation factor, the algorithm quantization unit $d_r = \frac{\lambda}{4} \frac{1}{N 2^n}$. By using $n = 7.3$ and $N = 250$ in the experimental demonstration with the optics described here, we obtain $d_r = \lambda/157\,600$, which is much higher resolution than the highest resolution obtained by the improved fringe-counting technique, which is approximately $\lambda/1000$. Therefore, quantization by the digital phase meter does not limit the resolution of the interferometer, which is instead limited by laser-frequency noise, shot noise, electrical noise, or other noise. To achieve high accuracy instead of high resolution, the periodic nonlinearity on the interferometer, which is normally induced by optical imperfection, should also be reduced. Optical techniques, such as spatially separating the source frequencies [29], can be applied in this situation.

For the figure of merit, we estimate the quantity of data $S$ per unit time used in the algorithm as follows: $S = 2^1(f_{\text{ADC}} - 4f_{\text{sig}}) + 2^n 4 f_{\text{sig}}$, where $f_{\text{sig}}$ is the frequency of the signal being measured. The first term on the right side of the equation shows that the information for data not involved in zero crossing is 1 bit. The second term shows that the information for the data involved in zero crossing is $n$ bits. For the conventional time-interval analysis, the quantity of data is simply expressed as $S = 2/t_{\text{TIA}}$, which reflects the fact that 1-bit data (positive of negative) is acquired with a time resolution $t_{\text{TIA}}$. By combining these two expressions for $S$, the effective time resolution of this algorithm is given as

$$t^{\text{eff}} = f_{\text{ADC}}^{-1}[1 + k(2^{n-1} - 1)]^{-1}, \qquad (2)$$

where the factor $k = 4f_{\text{sig}}/f_{\text{ADC}}$. Note that $k$ should be between 0 and 1 to satisfy the Nyquist condition. By using equation (2), the effective time resolution for three different ENOBs is plotted in Fig. 6. Here, the ENOBs are selected to be 7.3, 10.2, and 14, which corresponds to various commercially available digitizers. In addition, the sampling frequency of the ADC is fixed at 500 MHz. For the signal frequency of 80 MHz and the ENOB of 7.3 used in this experiment [see Fig. 3(c)], the effective time resolution is about 39.4 ps. This result shows that, with the proposed algorithm, the resolution is extended fifty-fold compared with the ADC sampling interval of 2 ns.

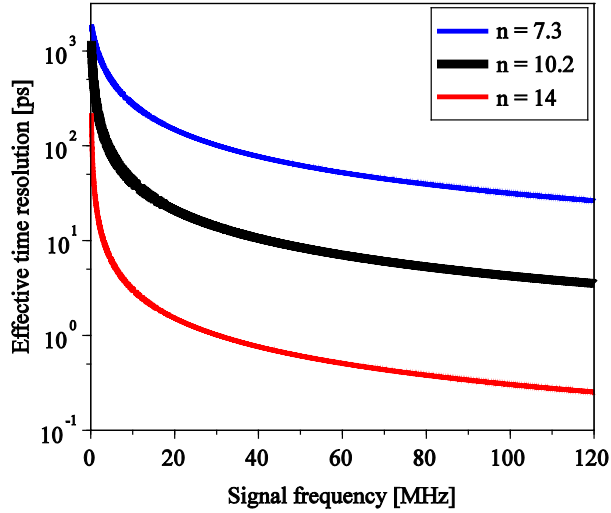

Fig. 6. Effective time resolution as a function of signal frequency for different effective number of bits of the ADC. The sampling frequency of the ADC is fixed at 500 MHz.

Another important feature of the proposed method is its ultrawide dynamic range. We compare this algorithm with the high-resolution lock-in method for heterodyne interferometry. With the lock-in method, the target velocity is limited by the DFT bandwidth. In particular, the maximum target velocity is $v_{\text{max}} = f_s \lambda/4$. If the result rate $f_s$ is 2 MHz, as in our demonstrative optical experiment, $v_{\text{max}} = 0.32$ m/s. In contrast, the proposed algorithm works when the modulated heterodyne signal satisfies the Nyquist condition for the ADC. In this case, we obtain $v_{\text{max}} =$ $(f_{\text{ADC}} - 4f_{\text{h}})\lambda/8$, where, $f_{\text{ADC}}$ and $f_{\text{h}}$ are the ADC sampling frequency (500 MHz in this experiment) and the heterodyne frequency (80 MHz), respectively. Therefore, $v_{\text{max}} = 14.2$ m/s when the proposed algorithm is adapted, extending the velocity range by approximately fiftyfold.

From the viewpoint of vibration metrology, the proposed phase meter has the advantage that it does not require an analog demodulator or an analog integrator (or LPF) because these components may result in additional uncertainties [30]. Considering that, in this investigation, we follow the national standards for low-frequency vibration [31] and shock [27], the phase measuring technique may be favorable in such systems. Other metrology experiments also require sensing dynamic displacement, including dynamic motion of a positioning stage [32], a Watt balance [33,34], dynamic pressure calibration [35], and absolute gravimetry [22].

Phase- and frequency-noise metrology may also benefit from this algorithm. In particular, it allows direct measurement of relatively large phase noise that exceeds the input dynamic range of conventional phase-noise measuring instruments because, in return for very low noise, the lock-in process used in digital phase-noise measurements [36] limits the dynamic range of phase or frequency. In addition, the direct time-domain phase information offers a much more sophisticated analysis of phase noise. Other than the phase-noise measurement, two periodical signals with very different frequencies can be directly compared in demodulator for frequency-modulated signals or in phase detectors for digital phase-locked-loop circuits. This feature may contribute to decreasing the lock-up time for such systems.

Birefringence measurements [37], accelerometers for disaster warning [38], and other precise measurement systems may also benefit from this technique because of its high resolution with respect to dynamic phase change. In addition, several space-based gravitational-wave detectors such as LISA [39] and DECIGO [40] and the GRACE follow-on [41] for observing Earth's gravitational field may take advantage of this technique for simple and precise displacement measurements in orbit.

## 5. Conclusion

In the present study, we propose an algorithm to estimate the phase of sinusoidal signals and that enables the use of a novel digital phase meter with reduced computation load. The algorithm is demonstrated in three experiments, and the noise levels obtained are comparable with those of other low-noise phase-measuring techniques that use lock-in amplifiers. Resolution and dynamic range in target velocity is improved fifty-folds with the proposed phasemeter. This type of dynamic phase-measuring technology can be applied in many fields in the industry and science to simplify wideband, high-resolution measurements. The present analysis is only based on the theoretical principles underlying the algorithm and on the experiments conducted to test its potential and applicability. A hardware implementation based on a FPGA is currently in process to enable real-time signal processing.

**Acknowledgments**

The authors acknowledge Keisuke Nakamura, Masato Wada, Sho Okubo, and Hajime Inaba for useful discussions. This work was financially supported by National Metrology Institute of Japan (NMIJ) and the National Institute of Advanced Industrial Science and Technology (AIST), Japan. The concept of the phase-measuring algorithm and its application including the heterodyne interferometer, phase-noise measuring instrument, and phase-locked loop is the subject of the patent application PCT/JP2015/081984.